\def \AAP #1 #2 {{\em Astron. Astrophys.\/} {\bf #1}, #2}
\def \AAL #1 #2 {{\em Astron. Astrophys. Lett.\/} {\bf #1}, L#2}
\def \AAR #1 #2 {{\em Astron. Astrophys. Rev.\/} {\bf #1}, #2}
\def \AAS #1 #2 {{\em Astron. Astrophys. Suppl. Ser.\/} {\bf #1}, #2}
\def \AJ #1 #2 {{\em Astron. J.\/} {\bf #1}, #2}
\def \ANNREV #1 #2 {{\em Ann. Rev. Astron. Astrophys.\/} {\bf #1}, #2}
\def \APJ #1 #2 {{\em Astrophys. J.\/} {\bf #1}, #2}
\def \APJL #1 #2 {{\em Astrophys. J. Lett.\/} {\bf #1}, L#2}
\def \APJS #1 #2 {{\em Astrophys. J. Suppl.\/} {\bf #1}, #2}
\def \APSS #1 #2 {{\em Astrophys. Space Sci.\/} {\bf #1}, #2}
\def \ASR #1 #2 {{\em Adv. Space Res.\/} {\bf #1}, #2}
\def \BAIC #1 #2 {{\em Bull. Astron. Inst. Czechosl.\/} {\bf #1}, #2}
\def \JSQRT #1 #2 {{\em J. Quant. Spectrosc. Radiat. Transfer\/} {\bf #1}, #2}
\def \MN #1 #2 {{\em Mon. Not. R. Astr. Soc.\/} {\bf #1}, #2}
\def \MEM #1 #2 {{\em Mem. R. Astr. Soc.\/} {\bf #1}, #2}
\def \PLR #1 #2 {{\em Phys. Lett. Rev.\/} {\bf #1}, #2}
\def \PASJ #1 #2 {{\em Publ. Astron. Soc. Japan\/} {\bf #1}, #2}
\def \PASP #1 #2 {{\em Publ. Astr. Soc. Pacific\/} {\bf #1}, #2}
\def \NAT #1 #2 {{\em Nature\/} {\bf #1}, #2}
\def \SAIT #1 #2 {{\em Mem.\ Soc.\ Astron.\ It.\/} {\bf #1}, #2}
\def \MESS #1 #2 {{\em The Messenger\/} {\bf #1}, #2}
\def \ASTRNACH #1 #2 {{\em Astron. Nach.\/} {\bf #1}, #2}
\documentstyle{BSAXPwork}\input epsf.sty\begin{opening}
\title{Accretion Models for Young Neutron Stars}
\author{M. Al\.{I} Alpar}
\institute{Sabanc{\i} University, Istanbul 81474,  Turkey}
\date{} 
\end{opening}

\begin{document}

\oddpagefooter{}{}{} 
\evenpagefooter{}{}{} 
\medskip  

\begin{abstract} 
Interaction with possible fallback material,  along with the magnetic fields and rotation rates at birth should 
determine the fates and categories of young neutron stars. This paper addresses some issues related to pure or hybrid accretion models for explaining 
the properties of young neutron stars.
\end{abstract}

\medskip

\section{AXP as single neutron stars}

Accretion from a fossil disk was first proposed by Van Paradijs, Taam \& Van den Heuvel (1995) to explain the X-ray 
luminosity in the apparent absence of a binary companion: a neutron star with a fossil disk would result from the final spiral-in destruction of the 
binary. The ages of such systems however is in conflict with the apparent youth of AXPs indicated by the SNR associations of some 
of them ( Gaensler 2001, Tagieva \& Ankay 2003). Accretion  from a fallback disk left from the core collapse was proposed as a model for AXPs 
by  Chatterjee, Hernquist \& Narayan (2000). In a simultaneous and independent paper (Alpar, 2001)  I proposed fallback disks as an option 
for all young neutron stars, indeed as a factor that must be one of the determinants of the fates of all young neutron stars. Marsden et al (2001) 
argued that the presence and effectiveness of fallback disks is correlated with SNR morphology and SNR/ISM environment, but the AXP-SGR and SNR 
correlations they claimed are not supported ( Gaensler 2001, Tagieva \& Ankay 2003).

The fallback disk idea goes back to earlier work on supernovae which suggested that matter, with angular momentum, 
can remain bound on the neutron star after the supernova explosion (Chevalier 1989,  Mineshige et al 1993). 
If so, the amount (or absence) of mass inflow  $ \dot{M}(t)$  towards/onto the neutron star, accretion, the propeller effect and the associated 
torques constitute one set of determinants, along with the magnetic dipole (and other multipole) moments as the other set of parameters, 
in setting the life path of the neutron star. Thus, independently of its relevance to AXPs and SGRs, the possibilities of fallback disks and 
their effects on the evolution and dynamics of young neutron stars has emerged as a basic problem of general interest.

One basic motivation for accretion/propeller models is that the narrow range of rotation periods, common to AXP and SGR, and also to the 
Dim Thermal Neutron Stars, has a natural explanation in terms of rotational equilibrium between neutron star and disk. Transitions between 
accreting and propeller states (transients- Cui et al 1998, Campana et al 1998), and between spinup and spindown 
(eg Lovelace, Romanova \& Bisnovatyi-Kogan 1999) occur 
near the equilibrium regime. As the rotational equilibrium is approached asymptotically, systems would tend to spend a long part of their lives 
near equilibrium and hence tend to be observed at their equilibrium periods. AXPs are spinning down while accreting because the fallback disk 
is evolving, with the inner disk radius receding from the neutron star, and the equlibrium period, which the star tracks, increasing as the mass 
inflow rate decreases (Chatterjee, Hernquist \& Narayan 2000).  The narrow range of observed periods correspond to equilibrium periods with 
dipole surface magnetic fields in the 10$^{12}$ G range and a not very restrictive range of mass inflow rates (Alpar 2001): AXP, SGR and DTN periods, 
with 10$^{12}$ G fields, imply an $ \dot{M}(t)$ range restricted to two orders of magnitude. There is a selection based on $ \dot{M}(t)$. 
According to the suggested classification AXPs and SGRs are the sources that get to equilibrium accretion before the disk depletes. 
DTNs  are also close to equilibrium but still in the propeller phase. With lower 
$ \dot{M}(t)$ , longer lives and slower evolution the DTNs are much more numerous than the younger AXPs and SGRs. Lowest or 
zero $ \dot{M}(t)$ give the radio pulsars, higher mass inflow rates all the way to Eddington give the 'radio quite' neutron stars (Alpar 2001). 
\\
\\
   
\section{Magnetars and Accretion}

Magnetar models provide a physical model for SGR bursts, energy output and super-Eddington fluxes. 
With the AXPs now being observed to experience bursts also (Gavriil, Kaspi \& Woods 2002) the magnetar models 
have gained in general applicability. In these models catastrophic release of energy from an 
intense magnetic field anchored in the neutron star provides the source of the bursts. Continuous energy 
dissipation as the magnetic field decays, rather than accretion, provides the source of X-ray luminosity of 
AXPs and SGRs in their quiescent phases. The detection of optical pulsations at a relatively large pulsed fraction from the AXP 4U 0142+61 
(Kern \& Martin 2002) also supports the magnetar models.

Neutron star spindown with magnetic dipole radiation at constant dipole moment cannot explain the narrow range 
of periods  (Psaltis \& Miller 2002). Magnetar models with field decay can explain the period clustering of AXPs and SGRs 
only under one particular class among the models for magnetic field decay inside the neutron star (Colpi, Geppert  \& Page 2000).  
The existence of 
some large magnetic field radio pulsars in the same region of the P-$ \dot{P}$ diagram as the AXPs and SGRs poses 
a number of questions as to the critical value of the surface dipole field at which the radio pulsar mechanism is 
quenched, and why, at field values that are quite close, the  magnetic decay mechanism producing the X-ray 
luminosity of AXPs and SGRs is not operative for the high field radio pulsars. If field decay  turns out to be the 
actual operating mechanism, the X-ray luminosity of AXPs and SGRs  will be recognized as restrictive and 
dependent on the range of field strengths, as a nonlinear process would. In magnetar models the selection, 
demanded by the rarity of the AXPs and SGRs, is provided by the selection of initial magnetic moments. 
In accretion models, the narrow range of periods finds an explanation in very general terms, of angular 
momentum equilibrium, while the selection of the observed classes of objects is provided by the range 
of mass inflow histories. Perhaps the AXPs and SGRs do have magnetar fields that provide the bursts, 
but the narrow range of periods is indeed due to rotational equilibrium with a fallback disk? This may be 
supported by the observation that AXPs and SGRs are closer to  accreting sources than to isolated radio 
pulsars in their timing properties (Alpar \& Baykal 2003).

Let us consider SGRs and AXPs in a hybrid model, with magnetar fields 
and fallback disks.  Pulsar magnetic fields show a distribution peaking strongly
 in the 10$^{12}$ G range with a tail extending well into the 10$^{13}$  G  range; There 
are radio pulsars with inferred surface dipole fields, at the pole, that exceed 
the quantum critical field, B$_{cr}$ = 4.4 10$^{13}$ G (Camilo et al 2000). Magnetars are 
in the tail end of this distribution as there is no indication that the B distribution 
is bi-modal. To retain the explanation of the observed range of rotation 
periods with magnetar fields, one must ask whether  the observed 
torques (spindown rates), luminosities of the neutron star and luminosities 
expected from the disk and the equilibrium periods can all be understood with 
a consistent and reasonable $ \dot{M}(t)$ range.
Torques exerted by the disk on the neutron star were estimated as:  

\begin{equation}
|\dot{\Omega}| =\mu^2 / r_{A}^3 (1- \Omega/\Omega_K(r_A))
\sim \mu^{2 / 7}  \dot{M}(t) ^{6 / 7}  [1- \Omega/\Omega_K(r_A)] 
\end{equation}
\\
\\
\\
\\
(Alpar 2001, Chatterjee, Hernquist \& Narayan 2000).  The inner radius of the disk depends on the rotation rate of the neutron star, 
$R_A(\Omega) > r_A $ (eg Lovelace, Romanova \& Bisnovatyi-Kogan 1995, Psaltis 2001). 
We adopt the estimates (Psaltis 2001) 

\begin{eqnarray}
R_A(\Omega)  \sim   r_A [1- \Omega/\Omega_K(r_A)]^{ -  2 / 7 } > r_A  \quad   \quad  (accretion) \nonumber \\
R_A(\Omega)  \sim  r_A [1- \Omega_K(r_A)/\Omega]^{ -  1 / 5 }  > r_A   \quad   \quad   (propeller).
\end{eqnarray}

For the accretion regime the dimensional torques are:

\begin{equation}
I |\dot{\Omega}| = \mu^2 / {r_A}^3 [1- \Omega/\Omega_K(r_A)]^{\;6 / 7}  \sim \mu^{2 / 7}  \dot{M}(t) ^{6 / 7}  [1- \Omega/\Omega_K(r_A)] ^{\;6 / 7} 
\end{equation}

The observed AXP and SGR spindown rates, with the $ \dot{M}(t)$ inferred from the quiescent L$_X$ and with magnetar values of the 
magnetic dipole moment, $\mu \sim 10^{32}$ G cm$ ^3$ require  $[1- \Omega/\Omega_K(r_A)]$ to be some 10$^{1/3} \sim $5 times smaller than 
the value needed with a conventional $\mu \sim10 ^{30}$ G cm$ ^3$ ? This is feasible, with $\Omega/\Omega_K(r_A) \sim $0.4 ,
consistent with the system being in an asymptotic regime.

Upper limits and detected luminosities in the optical and IR (Hulleman et al 2001, Kaplan et al 2001) restrict thin disk models. 
While these are the only available models, usefully applied by Chatterjee, Hernquist  \& Narayan (2000) to calculate AXP evolution 
scenarios with the fallback disk model, they are not necessarily realistic descriptions of a fallback disk, going through propeller phase, 
possibly with outflows, a corona, mass returning to disk, irradiation. These effects are likely to alter the spectrum of the disk significantly 
from the standard thin disk spectrum. Nevertheless, disk luminosities are not likely to differ much from the estimates of energy dissipation 
rates at the inner disk. Taking into account the dependence of the inner radius of the disk on $\Omega$, one arrives at 

\begin{equation}
L_{disk}   =  (GM \dot{M}(t) / r_A ) \, ( 1/2 \; [1- \Omega/\Omega_K(r_A)]^{ \;2 / 7 }  - \Omega/\Omega_K(r_A) [1- \Omega/\Omega_K(r_A)]^{\;6 / 7 }  )
\end{equation}

during accretion. Near equilibrium in the accretion regime, $\Omega/\Omega_K(r_A) \sim $ 1 , the disk luminosity is reduced in ratio to 
(GM $ \dot{M}(t) / r_A )$ . In the magnetar case, the expected disk luminosity scale $(GM  \dot{M}(t) / r_A )$ is lower, by a factor of 10 or 
more, than it is for conventional $\mu \sim $ 10 $^{30}$ G cm$ ^3$ since $r_A  \sim \mu ^{4 / 7} \dot{M}(t) ^{-2 / 7}$  is larger .

In the propeller regime the balance of the power expended by disk torques, after accounting for the neutron star spindown, goes into energy 
dissipation in the disk as well as powering any outflows. One finds 

\begin{eqnarray}
( L_{disk} + L_{outflow} ) = & ( & GM  \dot{M}(t) / r_A ) \, ( 1/2 \; [1- \Omega_K(r_A)/\Omega] ^{ \;1 / 5 } \nonumber \\
& + & \Omega/\Omega_K(r_A) [1- \Omega_K(r_A)/\Omega] ^{  \;3 / 5 })
\end{eqnarray}

Thus in the propeller regime also the disk luminosity is reduced, both through the  reduction in the luminosity scale (GM $\dot{M}(t)$ / r$_A$ ) for the 
magnetar case, and further if the system is  near equilibrium, $\Omega/\Omega_K(r_A)  \sim $ 1 .  
(There is, however, an enhanced luminosity in the extreme propeller regime, when $\Omega \gg \Omega_K(r_A)$ ). 

While extending the accretion model to a neutron star with magnetar fields is thus compatible with the observed torques and disk luminosities,  
the starting point of the accretion models, that the observed period clustering reflects the range of equilibrium periods cannot be retained for dipole 
fields in the magnetar range: $\Omega_{eq} = (GM / {r_A}  ^3 ) ^{1 / 2} \sim \mu ^{ - 6 / 7}  \dot{M}(t)^{ 3 / 7}$, so that the agreement 
with observed rotation rates using $ \dot{M}(t)$ indicated by the luminosities together with 
$\mu \sim $10$^{ 30}$ G cm$ ^3$  is lost for magnetars. For the same luminosities, magnetars would have equilibrium periods that are about 100 times 
larger than the observed range. A hybrid model will be possible only if the dipole magnetic moment is still $\mu \sim $10$^{ 30}$ G cm$ ^3$ while 
the magnetar field B $\sim $10$^{14}$ G on the neutron star surface has only higher multipoles.

\section{AXPs and Thin Disk Models}

Thin disk models are the only available models for the evolution of an isolated disk. Self similar solutions with a power law time evolution of 
the disk mass and mass inflow rate were employed for isolated disks (Canizzo et al 1990, Mineshige et al 1993) and were used by Chatterjee, 
Hernquist \& Narayan (2000) to model AXP evolution. The observations and upper limits in the optical and IR ranges 
(Hulleman et al 2001, Kaplan et al 2001) are in conflict with the 
disk luminosities predicted by the standard thin disk models, which are nevertheless usefully employed to provide a calculable model. Real 
fallback disks may not be thin diks at all, especially after a propeller phase,  or they may have comptonizing coronae.  At present there are no realistic 
models for comparison with the observations. Even within thin disk models the dependence of the disk inner radius on rotation rate 
will reduce the disk luminosity for systems near rotational equilibrium as we noted above. 

Francischelli \& Wijers (2002) have argued that a thin disk cannot spin the ns down to AXP periods within the AXP ages inferred from supernova 
remnant assocations unless  the magnetic field  B is larger than 3.7 $\times$ 10$^{13}$ G. Thin disk evolution with $ \dot{M}(t) \propto t ^{-\alpha}$
depends sensitively on opacity through the power law index $\alpha$. Francischelli \& Wijers note that Kramers opacities should prevail in the disk, 
and have followed the disk evolution 
with the corresponding power law index $\alpha$ = 1.25. However, they keep the same initial mass for the disk in their calculations with 
different power laws. For a given $\alpha$ the evolution may be insensitive to the initial mass in a certain range, but for different $\alpha$ there are
different choices of the initial disk mass which can lead to AXP formation. Furthermore the disk material is likely to be  iron rich 
(Fryer, Colgate \& Pinto 1999), in which case the opacities are not well known and will not lead to a generally valid power law index, so that the 
evolution must be solved numerically.

A self similar solution with power law time dependence holds for a disk that extends all the way to r = 0. There are two varieties of 
such self similar solutions of the thin disk diffusion equation. The type of solution discussed above involves power law evolution of the disk mass 
while the disk angular momentum J$_{disk}$ remains constant. The other type of self similar solution has constant disk mass while the disk angular 
momentum decays with  a  power law. The  latter type of solution may be the appropriate way to use a thin disk model to describe the propeller 
regime, (mass outflow returns to disk). In either case a real disk does not extend down to r = 0, but is cutoff at a finite inner radius r$_{in} (t)$. Starting 
off with power law evolution and using it at all times is not realistic. Numerical calculations for thin disk evolution with finite r$_{in}$ are found to settle to 
self-similar solutions with power law decay of both types, but only for appropriate intervals of the evolution. With all these considerations it is likely 
that a thin disk model may give spindown to AXP periods even with 10$^{12}$ G fields (Ek\c{s}i 2003).

\section{On Dim Thermal Neutron Stars }

DTNs are nearby, and therefore abundant objects, with ages $\sim$ 10$^6$ yrs. The classification in terms of fallback disk 
environments (Alpar 2001) interprets DTNs as slowly evolving objects, in the propeller phase in interaction with fallback disk of 
low mass and low mass inflow rate. The low luminosity is supplied by energy dissipation in the neutron star, in response to propeller 
spindown torques (unless the neturon star still has a cooling luminosity larger than the dissipation luminosity). 

Measured $\dot{\Omega}$ agrees with the dissipation luminosity due to dipole spindown in the case of RBS 1223 (Hambaryan et al 2002). 
For RXJ 0720.4-3125 the measured 
$\dot{\Omega}$ cannot supply the observed luminosity (Kaplan et al 2002, Zane et al 2002). In this case the luminosity must be due to cooling. This is 
reasonable as  standard cooling luminosities are high until about 10$^6$  yrs. 

If  DTNs are magnetars the energy dissipation driven by the magnetic dipole torque is 
much less than the cooling luminosity: L$_{diss}\sim$ 10$^{28}$ erg/s  t$_6^{ -3/2} \mu_{32}^{ -1}$ . In magnetar models the luminosity of DTNs would be 
due to cooling with a possible contribution supplied by field decay. For RXJ 0720.4-3125 bounds on the spindown rate rules out a magnetar 
(Kaplan et al 2002). In evaluating magnetar models for the period clustering of AXPs Colpi, Geppert \& Page (2000) found that only one class of magnetar 
field decay models, Hall cascade in the crust of the neutron star, is consistent with the observed period clustering, luminosities and inferred ages of 
the AXPs, if the initial magnetic field strength is 10$^{15}$ G. But this scenario extrapolates to luminosities of 10$^{33}$ erg/s at the 10$^6$ yr ages of 
the DTNs (Geppert et al 1999), higher than the observed DTN luminosities by an order of magnitude at least! 

Could all the DTN luminosities be just due 
to the the cooling of an isolated neutron star with a conventional magnetic field? 
DTNs seem to cluster in luminosity at about 10$^{32}$  ergs / s . If this is borne out by more statistics, and the luminosity is due to cooling, then they 
cluster in age at about 10$^6$ yrs! Why are there no older ones then? This is difficult to explain in terms of cooling luminosity.

If the luminosity is supplied by energy dissipation for most DTNs (those stars that are past the neutrino cooling luminosity era) its value will indeed stay 
roughly constant at  L$_{diss}$ = 10$^{32}$ erg/s for the duration of the propeller phase close to rotational equilibrium. Thus the age and luminosity 
(distance) determinations of the DTNs hold the clue as to whether they are propellers or cooling neutron stars under dipole spindown.

\acknowledgements

\end{document}